\documentclass[angeocom]{copernicus}

\begin{document}

\title{\hack{\vspace{3mm}}Observation of a new type of low-frequency waves at comet 67P/Churyumov-Gerasimenko}

\Author[1]{I.}{Richter}
\Author[1]{C.}{Koenders}
\Author[1]{H.-U.}{Auster}
\Author[1]{D.}{Fr\"uhauff}
\Author[1]{C.}{G\"otz}
\Author[1]{P.}{Heinisch}
\Author[1,2]{C.}{Perschke}
\Author[2,3]{U.}{Motschmann}
\Author[1]{B.}{Stoll}
\Author[4]{K.}{Altwegg}
\Author[5]{J.}{Burch}
\Author[6]{C.}{Carr}
\Author[6]{E.}{Cupido}
\Author[7]{A.}{Eriksson}
\Author[8]{P.}{Henri}
\Author[5]{R.}{Goldstein}
\Author[8]{J.-P.}{Lebreton}
\Author[5]{P.}{Mokashi}
\Author[9]{Z.}{Nemeth}
\Author[10]{H.}{Nilsson}
\Author[4]{M.}{Rubin}
\Author[9]{K.}{Szeg\"o}
\Author[11]{B. T.}{Tsurutani}
\Author[12]{C.}{Vallat}
\Author[13]{M.}{Volwerk}
\Author[1]{K.-H.}{Glassmeier}

\affil[1]{Institut f\"ur Geophysik und extraterrestrische Physik, TU Braunschweig, Mendelssohnstr. 3, \hack{\newline}  38106 Braunschweig, Germany}
\affil[2]{Institut f\"ur Theoretische Physik, TU Braunschweig, Mendelssohnstr. 3, 38106 Braunschweig, Germany}
\affil[3]{German Aerospace Center (DLR), Institute of Planetary Research, Rutherfordstra\ss{}e 2, \hack{\newline}  12489 Berlin, Germany}
\affil[4]{Physikalisches Institut, University of Bern, Sidlerstrasse 5,  3012 Bern, Switzerland}
\affil[5]{Southwest Research Institute, P.O.~Drawer 28510, San Antonio, TX 78228-0510, USA}
\affil[6]{Imperial College London, Exhibition Road, London SW7 2AZ, UK}
\affil[7]{Swedish Institute of Space Physics, {\AA}ngstr\"{o}m Laboratory, L\"{a}gerhyddsv\"{a}gen 1, Uppsala, Sweden}
\affil[8]{Laboratoire de Physique et Chimie de l'Environnement et de l'Espace, UMR 7328 CNRS -- \hack{\newline}  Universit\'{e} d'Orl\'{e}ans, Orl\'{e}ans, France}
\affil[9]{Wigner Research Centre for Physics, 1121 Konkoly Thege Street 29--33, Budapest, Hungary}
\affil[10]{Swedish Institute of Space Physics, P.O.~Box 812, 981 28 Kiruna, Sweden}
\affil[11]{Jet Propulsion Laboratory, California Institute of Technology, 4800 Oak Grove Drive, Pasadena, \hack{\newline}  CA 91109, USA}
\affil[12]{Rosetta Science Ground Segment, European Space Astronomy Centre, 28691 Villanueva de la Ca\~{n}ada, \hack{\newline}   Madrid, Spain}
\affil[13]{Space Research Institute, Austrian Academy of Sciences, Schmiedlstra{\ss}e 6, 8042 Graz, Austria}

\runningtitle{Low-frequency waves at comet 67P/Churyumov-Gerasimenko}

\runningauthor{I.~Richter et al.}

\correspondence{I.~Richter (i.richter@tu-braunschweig.de)}

\received{22 May 2015}
\revised{14 July 2015}
\accepted{17 July 2015}
\published{19 August 2015}

\firstpage{1}

\maketitle

\begin{abstract}
We report on magnetic field measurements made in the innermost coma of
67P/Churyumov-Gerasimenko in its low-activity state. Quasi-coherent,
large-amplitude ($\delta B/B \sim 1$), compressional magnetic field
oscillations at $\sim$\,40\,mHz dominate the immediate plasma environment of
the nucleus. This differs from previously studied cometary interaction regions
where waves at the cometary ion gyro-frequencies are the main feature. Thus
classical pickup-ion-driven instabilities are unable to explain the
observations. We propose a cross-field current instability associated with
newborn cometary ion currents as a possible source mechanism.
\end{abstract}

\introduction
Typically ionization of atoms and molecules of cometary origin is the most
important process for the interaction of strongly outgassing comets and the
solar wind. During encounters of the ICE, Sakigake, and Giotto spacecraft
with active comets 21P/Giacobini-Zinner, 1P/Halley, and 26P/Grigg-Skjellerup,
large-amplitude plasma waves and turbulence have been one of the most
pronounced observational findings in the cometary magnetosphere
\citep{Tsurutani1986a,Yumoto1986,Neubauer1986,Glassmeier1989,Glassmeier1993a,Volwerk2014}. Ion ring-beam instabilities \citep{Wu1972} and non-gyrotropic
phase space density-driven instabilities \citep{Motschmann1993} are the
source mechanism of these waves. In the spacecraft (s/c) frame of reference (and pickup
ion frame), those waves were detected at the cometary \chem{H_2O^+} ion
gyro-frequency.

Rosetta's journey \citep{Glassmeier2007a} alongside comet
67P/Churyumov-Gerasimenko now allows for electro-magnetic waves to be studied at the
beginning of cometary activity, at the birth of the cometary magnetosphere
\citep{Nilsson2015}. It should be noted that, under these low-activity
conditions, typical solar wind--cometary interaction regions like the bow shock
and magnetic pileup region are not expected (e.g., \citealp{Koenders2013};
\citealp{Rubin2014b}) and also not observed \citep{Nilsson2015}. We shall
report on wave observations at distances of 2.7--3.6\,AU from the Sun and
10--1000\,km from the comet.

\section{Mission and instrumentation}

Rosetta arrived at 67P/Churyumov-Gerasimenko on 6 August  2014 at a
heliocentric distance of 3.6\,AU. The spacecraft was initially put into a
$\sim$\,100\,km orbit around the comet's nucleus. Observations reported here are
limited to the dayside inner coma and sampled over a time span of a $\sim$\,4~months (from August to November 2014). Using measurements of the Rosetta
Orbiter Spectrometer for Ion and Neutral Analysis Cometary Pressure Sensor
(ROSINA COPS) \citep{Balsiger2007}, the cometary activity at this heliocentric
distance was determined to be below $4 \times 10^{26}$\,s$^{-1}$. This
production rate is 2--3 orders of magnitude lower than at any other previous
cometary encounter where pickup ion waves were detected ($8 \times 10^{29}$\,s$^{-1}$ at 1P/Halley, down to $7 \times 10^{27}$\,s$^{-1}$ at
26P/Grigg-Skjellerup) \citep[see][]{Richter2011}.

The Rosetta orbiter is equipped with a suite of plasma instruments -- the
Rosetta Plasma Consortium (RPC) set of particle and field sensors
\citep{Carr2007}. RPC-MAG, the tri-axial fluxgate magnetometer system
\citep{Glassmeier2007b}, consists of two sensors mounted on a 1.5\,m boom,
separated by 0.15\,m. The short boom length implies that the spacecraft is
heavily contaminating the magnetic field measurements. At this stage of the
investigation it was not possible to completely remove these quasi-static
spacecraft bias fields from the measured magnetic field values. The dynamic
range of RPC-MAG is $\pm$16\,000\,nT, and its resolution 0.03\,nT. Although the
magnetometer is capable of acquiring the magnetic field measurements with
sampling rates up to 20\,Hz, the data presented here correspond to the
instrument's normal operational mode, i.e., a sampling rate of 1\,Hz, which is
sufficient for the purposes of our investigation. The magnetic field
observations are represented in a comet-centered solar equatorial (CSEQ)
coordinate system. The $+x$~axis points from the comet to the Sun, the $+z$~axis
is the component of the Sun's north pole of date orthogonal to the $+x$~axis,
and the $+y$~axis completes the right-handed reference frame. The origin of the
coordinate system is the comet's center of mass.

The Rosetta Ion and Electron Sensor (RPC-IES) \citep{Burch2006} and the Ion
Composition Analyser (RPC-ICA) \citep{Nilsson2006} provide information on
cometary ions produced in the coma of 67P/Churyumov-Gerasimenko. Observations
of the neutral gas number density made by ROSINA COPS are used as
well.

\section{Observations}
Upon arrival at the comet on 6 August  2014, RPC-MAG started to detect
large-amplitude, quasi-coherent magnetic field fluctuations. During the
aforementioned observational period, we were able to collect $\sim$\,3000
cases of wave activity with typical frequencies of $\sim$\,40\,mHz. Figure~1
shows an example of these waves. The wave activity is clearly visible in all
three components. Peak-to-peak amplitudes are of the order of 4\,nT, which is
about twice as large as the ambient solar wind magnetic field at this
heliocentric distance. The oscillations are neither purely transverse nor
purely compressional.

A preliminary minimum variance analysis between August and November 2014 does
not show any preferred direction of wave propagation, with respect to neither
the solar wind flow nor the mean magnetic field direction. A
full discussion of the minimum variance analysis results will be presented
elsewhere. As only single point observations are available, because only the
ROSETTA orbiter magnetometer was operating and the lander was still attached
to the orbiter, no information about a typical wavelength can be inferred at
this time.

%f1
\begin{figure}[t]
\includegraphics[width=8.3cm]{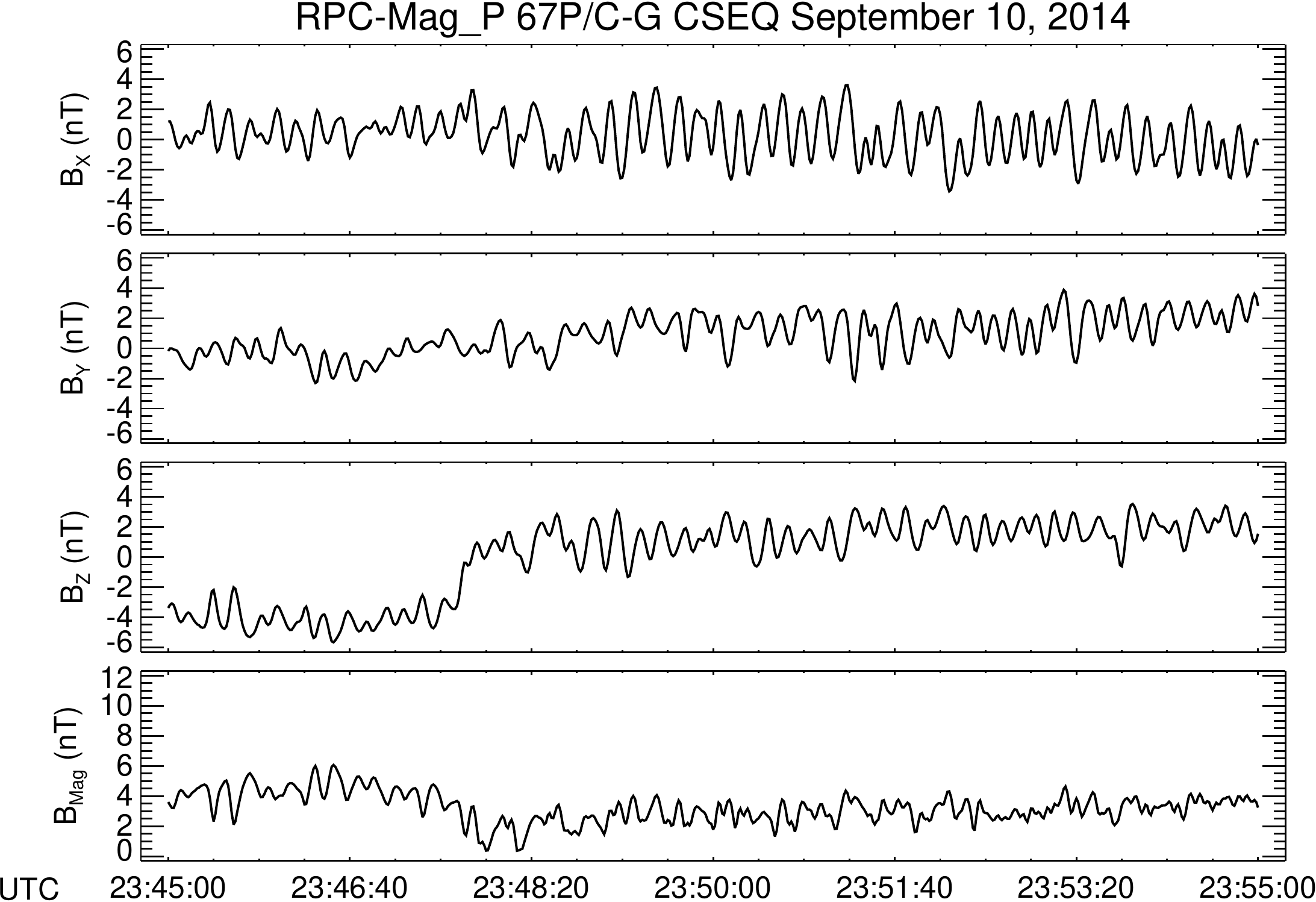}
\caption{Example of magnetic field observations made onboard the Rosetta
spacecraft on 10 September  2014, 23:45--23:55\,UTC. The position vector of the
spacecraft in the comet-centered solar equatorial (CSEQ; for details see text)
coordinate system was $(3.9,-20.6,20.4)$\,km.  \label{figure_1}}
\end{figure}

\begin{figure}[t]
\includegraphics[width=8.3cm]{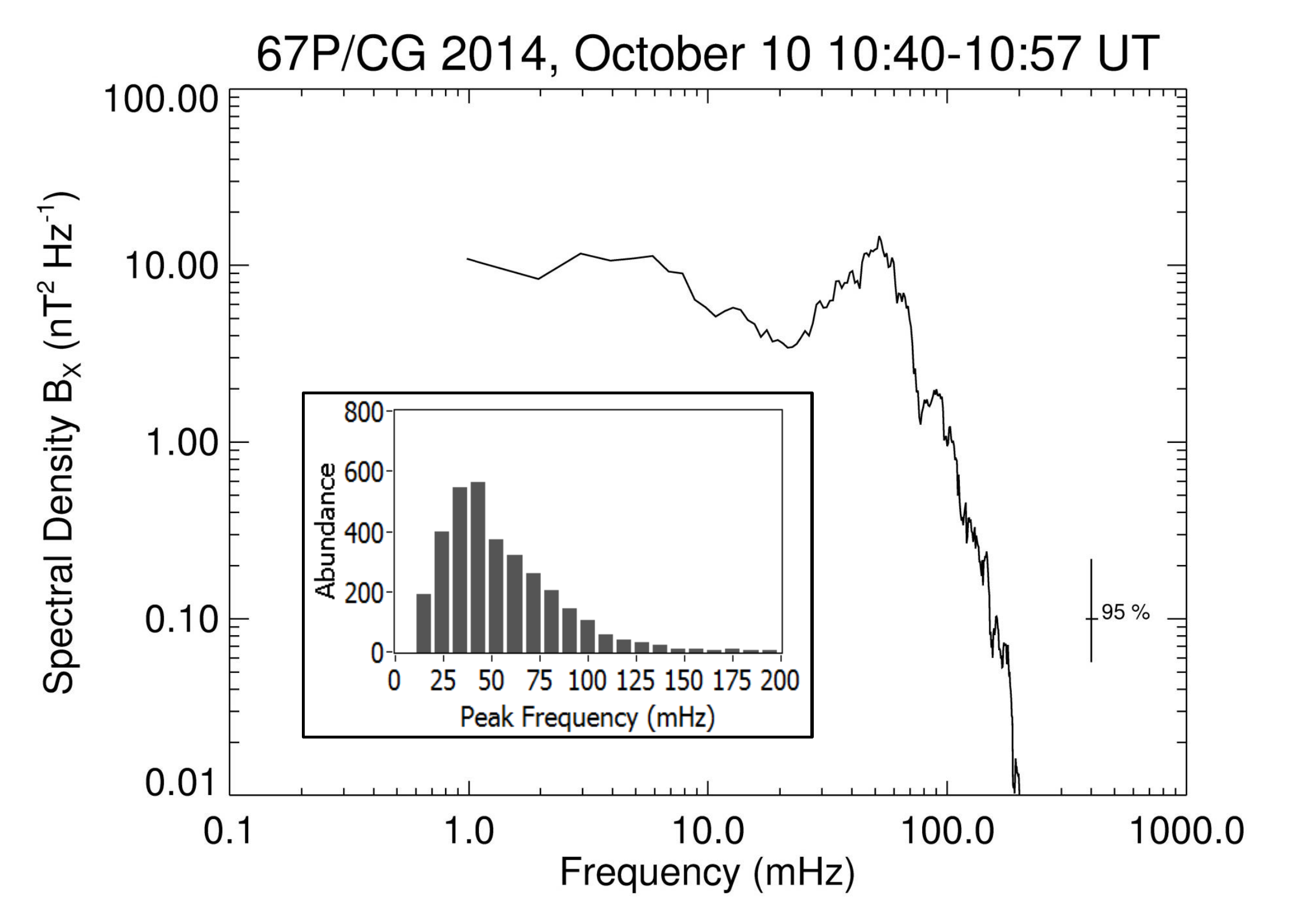}
\caption{Example of a spectrum of magnetic field fluctuations observed in the
innermost coma of 67P/Churyumov-Gerasimenko on 10  October 2014,
10:40--10:57\,UTC (CSEQ position vector $(-0.3,-10.1,-1.7)$\,km). The power spectrum has
been calculated by application of a standard fast Fourier transform routine to a 17\,min time
interval. A box car window is used to average in the frequency domain. Eighteen degrees of freedom are used. The inset shows the distribution of peak
spectral frequencies in the period August--November 2014. The confidence
interval is shown in the bottom right corner. \label{figure_2}}
\end{figure}

The quasi-coherent nature of these fluctuations is also clearly visible in
power spectral density distributions (Fig.~\ref{figure_2}). The steep
spectral slope at frequencies beyond the peak frequency is not uncommon for
the observations at 67P/Churyumov-Gerasimenko; typical spectral slopes are
between $-$3 and $-$5. The wave spectra typically exhibit a single-peak structure
as shown in Fig.~\ref{figure_2}.

%f2

The peak frequencies exhibit a Rayleigh-type distribution grouped  around 40\,mHz (Fig.~\ref{figure_2}).
Further statistical analyses do not show any clear correlation of the peak
frequencies with the ambient magnetic field magnitude. Such a correlation
would be expected if the observed frequency coincides with the local proton
gyro-frequency $f_\mathrm{p}$ ($f_\mathrm{p} \sim 40$\,mHz with $B=2.5$\,nT). Though the agreement between the peak frequency of the Rayleigh-type
distribution and the local gyro-frequency is striking, we argue here that
the observed waves are not in proton cyclotron resonance as a clear
correlation between magnetic field magnitude and frequency is missing.

\hack{\newpage}

\begin{figure}[t]
\includegraphics[width=8.3cm]{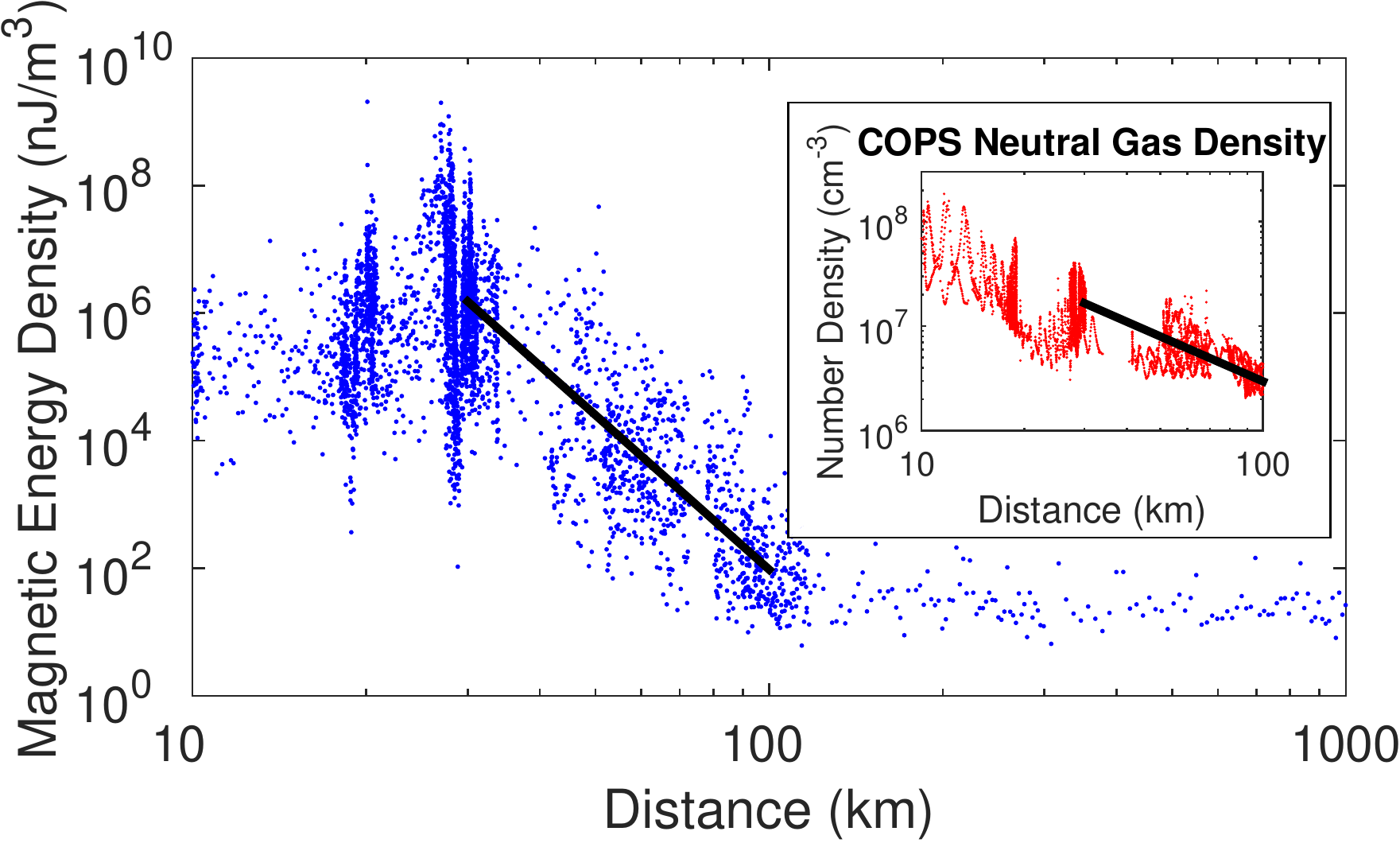}
\caption{Radial
variation in the magnetic energy density of the low-frequency wave activity
in the spectral range 30--80\,mHz. The inset shows the radial variation in the
density of cometary neutrals as measured by the neutral gas monitor COPS of
the ROSINA instrument.  \label{figure_3}}
\end{figure}

Wave activity was first observed by the RPC-MAG instrument at a dayside
distance of approximately 100\,km from the nucleus and steadily increased up
to a distance of 30\,km (Fig.~\ref{figure_3}). Hourly trace spectral densities
have been integrated over the frequency range 30--80\,mHz and divided by
twice the vacuum permeability to determine the magnetic energy density of the
observed fluctuations. Closer to the nucleus the magnetic energy density
seems to saturate. The spread of the distribution at around 20  and 30\,km
is caused by Rosetta's trajectory with respect to the comet. During the first
months after arriving at the comet the spacecraft was often been positioned
in the so-called bound orbits with a fixed distance to the nucleus, mostly in
the terminator plane \citep{Hassig2015}. Hence, more observations are made at
these distances. The wave activity variations at constant distance are caused
by temporal variations in the neutral gas density produced by localized gas
sources on the nucleus' surface and modulated by the comet's rotation
\citep{Hassig2015}, leading through the ionization of cometary neutrals to
temporal variations in the cometary plasma density. The variations also
reflect a dependence on solar wind variations as well as elevation and
azimuth angle of Rosetta's position. Future work is planned to separate these
dependencies. For the range of 30--100\,km the spatial variation in the magnetic
energy density is found to be $W \propto r^{-\alpha}$ with $\alpha
=8.04$\,$\pm$\,0.27 (for a 95\,{\%} confidence interval $\alpha \in
\left[7.77; 8.31\right]$, yielding a relative error of $100 \cdot \frac{\delta
\alpha}{\alpha}=3.3$\,{\%} ), equivalent to a quartic decrease in the wave
amplitude with distance.

Comparing wave intensity with the neutral gas number density as measured by
ROSINA COPS exhibits a clear global relation between both quantities. COPS also
detected a neutral gas number density, $N$, above its noise level at a
distance of about 100\,km (Fig.~\ref{figure_3}). The neutral density decreases
as $N \propto r^{-\beta}$ with $\beta =1.43$\,$\pm$\,0.07  (for a 95\,{\%}
confidence interval $\beta \in \left[1.36; 1.50\right]$). The deviation
$100 \cdot \frac{\beta_\mathrm{theo} - \beta_\mathrm{calc}}{{\beta_\mathrm{theo}}} =28.5$\,{\%} from the theoretically
expected value $\beta_\mathrm{theo}=2$ \citep{Coates2009} is
due to Rosetta first approaching from the afternoon side (45{\degree} phase
angle) before moving to the terminator plane (90{\degree} phase angle) and
therefore encountering different insolation conditions at the sub-spacecraft
location.

Here only a global relation is discussed. More detailed analyses on the
relation between neutral gas density and magnetic field variations are
currently being prepared; however, such studies require consideration of the rotation of
the nucleus with a period of about 12.4\,h as well as any inhomogeneity in
the active regions.

%f3

\conclusions[Discussion and possible wave source mechanism]
Comparing the radial variations in both the magnetic energy density, $W$, and the
neutral gas density suggests an approximate global relation $W \propto
N^{-\gamma}$ with $\gamma =  5.62 $\,$\pm$\,0.45  (for a 95\,{\%} confidence interval
$\gamma \in \left[5.17; 6.07\right]$). A more detailed correlation analysis
between the individual values of both quantities at the same radial distance
reveals a linear Pearson correlation coefficient of $r=0.52$. This correlation
reflects the large spread of both wave energy and neutral gas density. As
waves are generated locally under time-varying solar wind conditions and
propagate within the inner coma, this weak correlation is not surprising.
Because the production rate of heavy cometary ions is proportional to the
neutral gas density, we conclude that wave activity in general is controlled
by the cometary ion production rate.

The observed frequencies concentrate between 25 and 75\,mHz (Fig.~\ref{figure_2}).
In the coma the local H$_2$O$^+$ ion gyro-frequency is in the range of
 0.8--3.2\,mHz. Thus, there are clear differences between the observed
frequencies at the birth stage of the cometary magnetosphere and the heavy
ion cyclotron frequencies, as expected for well-developed cometary interaction
regions. This points towards a new generation mechanism for the type of wave
reported here for a weakly outgassing comet.

The size of the innermost interaction region is much less than the Larmor
radius of the newborn ions. The pickup ion's initial velocity is about 0.8\,km\,s$^{-1}$; acceleration by the interplanetary electric
field, $\vec E$, up to velocities, $\mid \vec v \mid$, of a few tens of kilometers per second is observed \citep{Nilsson2015} by RPC-IES
and RPC-ICA. After ionization the newborn ions are moving transverse to the
ambient magnetic field, $\vec B$, and the solar wind flow in the direction of
the electric field, constituting a cross-field electric current density. The
cometary ion motion is controlled by the electric field. Lorentz forces are
not yet important in this innermost coma region. The ratio of the electric
force to the Lorentz force is of the order of the ratio of the gyro-period
$T_\mathrm{G}$ to the lifetime $\tau$ of the newborn ions: $ \mid \vec E
\mid / \mid \vec v \times \vec B \mid \approx T_\mathrm{G}/\tau$. Newborn
cometary ions with $\tau \ll T_\mathrm{G}$ are essentially unmagnetized.
Lorentz forces become important only on scales comparable to and larger than a
cometary ion Larmor radius. Within distances to the nucleus smaller than a
Larmor radius, i.e., in the Larmor sphere \citep{Sauer1998}, physical
processes different from those generating the classical pickup-ion-related
ion cyclo\-tron waves are important.

A newborn water ion flux, $n \cdot v$, of at least $10^{10}$\,m$^{-2}$\,s$^{-1}$ has been observed by the RPC-ICA sensor
\citep{Nilsson2015}. Due to the sensor's limited field of view, the actual flux will be
higher. Assuming a flux density $3\times 10^{10}$\,m$^{-2}$\,s$^{-1}$ and singly charged ions the electric current density is
estimated to be about $j \sim 4.8 \times 10^{-9}\, \mathrm{A}$\,m$^{-2}$. Electromagnetic instabilities attributable to such a
cross-field current have been studied in the past \citep{Chang1990,Sauer1998}, but not for conditions typical for the plasma situation at
67P/Churyumov-Gerasimenko. Assuming that the cross-field current is driven
unstable, a transverse wave number $k_\perp$ can be estimated using
Ampe\`{e}re's law: $k_\perp = \mu_0 \delta j / \delta B$. With $\delta j \sim
j$ and $\delta B \approx 1$\,nT, a value of $k_\perp \approx
6\times 10^{-6}\,\mathrm{m}^{-1}$ results, corresponding to a transverse wave
length of 524\,km. It should be noted that this wave number component is
transverse to the mean magnetic field as well as the cross-field current. The
polarization of the observed waves is neither purely compressional nor purely
transverse. This points towards off-angle propagation. We therefore assume
that all three components of the wave vector are of comparable magnitude.

The wavelength reported here is larger than the scale of the dayside inner
coma where wave activity has been observed. However, the generation region is
probably significantly larger. But as waves generated trough cometary interaction at larger distances are buried in the pre-existing solar wind turbulence, the
signal was not detected by RPC-MAG. We need to await a further increase in
the cometary activity to see a further expansion of that region where
cometary waves clearly stand out from the solar wind turbulence
background.

The frequency of the unstable mode may be estimated using the beam-mode
dispersion relation \citep{Chang1990,Sauer1998} $\omega \approx k_\parallel
\cdot v$ , where $k_\parallel$ is the beam parallel wave vector component and
$v$ the cometary ion beam velocity.

Using an observational estimate \citep{Nilsson2015} of the ion velocity $v =
40$\,km\,s$^{-1}$, we obtain an angular frequency omega
$\omega \sim 0.24\; \mathrm{rad}\,\mathrm{s}^{-1}$, i.e., a wave frequency $f
\sim\, 38$\,mHz.

This value is comparable to the observed frequencies.

It should be noted that the suggested wave source is not co-moving with the
solar wind flow. The cross-field current source is due to freshly ionized
cometary ions which are not yet moving with the solar wind velocity as they
did not have time to be accelerated to solar wind speed on the time and
distance scales we are looking at. The wave source is almost fixed in the
nucleus frame of reference. Therefore, Doppler effects can be neglected as
the Rosetta spacecraft is only slowly moving with respect to the nucleus
($v_\mathrm{s/c} \sim 1$\,m\,s$^{-1}$).

As the wave number is proportional to the current density perturbation, $k
\propto \delta j \propto n\,v$, the dispersion relation provides the
following approximate expression between wave frequency, ion density and ion
velocity: $f \propto n v^2$. Assuming a constant electric field accelerating
the particles, the ion velocity should increase linearly with distance. For an
ion density decreasing with the square of the distance, the frequencies
should therefore not exhibit any major dependence on distance, which is what
we observe. Of course, this is only a first conjecture stimulating future
analysis.

Our model furthermore allows for the apparent saturation of wave
activity at around 30\,km distance to be understood: the newborn ions need to be accelerated to
constitute a significant current and the waves need to grow. However, further
detailed theoretical modeling is required to validate the conjecture of a
cross-field current-driven instability causing the newly detected
low-frequency wave activity in the Larmor sphere of
67P/Churyumov-Gerasimenko.

\begin{acknowledgements}
The RPC-MAG and ROSINA data will be made available through the PSA archive of
ESA and the PDS archive of NASA. Rosetta is a European Space Agency (ESA)
mission with contributions from its member states and the National
Aeronautics and Space Administration (NASA). The work on RPC-MAG was
financially supported by the German Ministerium f\"{u}r Wirtschaft und Energie
and the Deutsches Zentrum f\"{u}r Luft- und Raumfahrt under contract 50QP 1401.
The work on ROSINA was funded by the federal state of Bern, the Swiss National
Science Foundation, and the ESA PRODEX program. Portions
of this research were performed at the Jet Propulsion Laboratory, California
Institute of Technology, under contract with NASA. We are indebted to the
whole of the Rosetta Mission Team, SGS, and RMOC for their outstanding efforts in
making this mission possible.\hack{\newline}
\hack{\hspace*{4mm}} The topical editor  E.~Roussos thanks the two anonymous referees for help in evaluating this paper.
\end{acknowledgements}

\end{document}